\documentclass[useAMS,usenatbib,usegraphicx]{mn2e}
\usepackage{epsf}
\usepackage{amssymb,amsmath}

\title[Testing the cosmic curvature at high redshifts]{Testing the cosmic curvature at high redshifts: the combination of LSST strong lensing systems and quasars as new standard candles}

\author[Liu et al.]
{Tonghua Liu$^{1}$, Shuo Cao$^{1\star}$, Jia Zhang$^{2}$, Marek Biesiada$^{1,3\dagger}$, Yuting Liu$^{1}$, Yujie Lian$^{1}$  \\
$^1$ Department of Astronomy, Beijing Normal University, 100875, Beijing, China; \emph{caoshuo@bnu.edu.cn}\\
$^2$ School of physics and Electrical Engineering,
Weinan Normal University, Shanxi 714099, China;\\
$^3$ National Centre for Nuclear Research, Pasteura 7, 02-093
Warsaw, Poland; \emph{Marek.Biesiada@ncbj.gov.pl}}

\begin{document}

\date{\today}

\voffset- .5in

\pagerange{\pageref{firstpage}--\pageref{lastpage}} \pubyear{}

\maketitle

\label{firstpage}

\begin{abstract}

The cosmic curvature, a fundamental parameter for cosmology could
hold deep clues to inflation and cosmic origins. We propose an
improved model-independent method to constrain the cosmic curvature
by combining the constructed Hubble diagram of high-redshift quasars
with galactic-scale strong lensing systems expected to be seen by
the forthcoming LSST survey. More specifically, the most recent
quasar data are used as a new type of standard candles in the range
$0.036<z<5.100$, whose luminosity distances can be directly derived
from the non-linear relation between X-ray and UV luminosities.
Compared with other methods, the proposed one involving the quasar
data achieves constraints with higher precision ($\Delta
\Omega_k\sim 10^{-2}$) at high redshifts ($z\sim 5.0$). We also
investigate the influence of lens mass distribution in the framework
of three types of lens models extensively used in strong lensing
studies (SIS model, power-law spherical model, and extended
power-law lens model), finding the strong correlation between the
cosmic curvature and the lens model parameters. When the power-law
mass density profile is assumed, the most stringent constraint on
the cosmic curvature $\Omega_k$ can be obtained. Therefore, the
issue of mass density profile in the early-type galaxies is still a
critical one that needs to be investigated further.
\end{abstract}


\section{Introduction}\label{sec:introduction}

The cosmic curvature is one of the fundamental issues in modern
cosmology, which determines the evolution and structure of our
Universe. Specifically, the spatial properties of our universe is
not only closely connected with many important problems such as the
properties of dark energy (DE) \citep{Clarkson2007,Gong2007}, but
also influences our understanding of inflationary models
\citep{Ichikawa2006,Virey2008}, the most popular theories describing
the evolution of early universe. More importantly, any detection of
nonzero spatial curvature ($\Omega_k\neq0$) would have significant
consequences on the well-known FLRW metric, which has been
investigated in many recent studies \citep{Denissenya2018,Cao19a}.
Therefore, precise measurements of spatial curvature allowing to
better understand this degeneracy will have far-reaching
consequences. The recent Planck 2018 results imposed very strong
constraints on the curvature parameter, $\Omega_k=0.001\pm0.002$,
based on cosmic microwave background (CMB) anisotropy measurements
\citep{Planck Collaboration}. However, it should be stressed here
that the curvature inferred from CMB anisotropy data is obtained by
assuming some specific dark energy model (the non-flat $\Lambda$CDM
model). Therefore, it is necessary to consider different geometrical
methods to derive model-independent measurements of the spatial
curvature.

Following this direction, great efforts have been made in the recent
studies \citep{Cai2016,Li2016c,Wei2017,Wang2017,Rana2017}, with the
combination of the well-known cosmic chronometers (which provide the
expansion rate of the Universe \citep{Clarkson2008}) and the
observations of supernovae Ia (SNe Ia) (which provide the luminosity
distances at different redshifts \citep{Suzuki12}). Later, such test
has been implemented with updated observations of
intermediate-luminosity radio quasars \citep{Qi2019}, the angular
sizes of which could provide a new type of standard rulers at higher
redshifts \citep{Cao19b}. Recently, \citet{Rasanen14} proposed a
model-independent way to obtain constraints on the curvature, with
strong gravitational lensing (SGL) data in the framework of the
distance sum rule (DSR) in the FLRW metric
\citep{Takada2015,Denissenya2018,Ooba2018}. In the framework of
strong gravitational lensing (SGL) \citep{Cao2013,Cao2015a}, the
light can be bent by the gravity of massive body (at redshift
$z_l$), which could produce multiple images for the distant sources
(at redhsift $z_s$). Supplemented with the observations of the lens
central velocity dispersion, the Einstein radius measurement
\citep{Bolton08,Cao2012,Cao2015b} will enable a precise
determination of the source-lens/lens distance ratio $d_{ls}/d_s$
\citep{Cao11,Cao12a,Cao12b} for individual strong lensing system. In
addition, one should also estimate distances at redshifts $z_l$ and
$z_s$ from different astrophysical probes covering these redshifts
such as SNe Ia or Hubble parameters from cosmic chronometers
\citep{Clarkson2008,Clarkson2007,Shafieloo2010,Li2016c}. The
advantage of this method is that it is purely geometrical and the
curvature can be constrained directly by observational data, without
any pre-assumptions concerning the cosmological model and the FLRW
metric \citep{Cao19a}. Such methodology has been first implemented
with a SGL subsample from the Sloan Lens ACS Survey (SLACS)
\citep{Bolton08}, which favors a spatially closed universe with the
final results that the spatial curvature parameter could be
constrained to $-1.22< \Omega_k <0.63$ (95\% C.L.)
\citep{Rasanen2015}. More recently, several other studies have been
carried out with enlarged galactic-scale SGL sample
\citep{Cao2015b}, as well as updated observations of SNe Ia data and
radio quasars as distance indicators \citep{Cao17a,Cao17b}, which
furthermore confirmed the robustness of such consistency test as a
practical measurement of the cosmic curvature
\citep{Xia2017,Qi2019,Zhou20}. For instance, it has been
demonstrated in a recent analysis \citep{Qi2019} that 120
intermediate-luminosity radio quasars calibrated as standard rulers
($z\sim2.76$), in combination with 118 galactic-scale strong lensing
systems, could provide an improved constraint on cosmic curvature
$\Omega_k<0.136$. However, it should be pointed out that, the
previous results still suffer from the sample size of available SGL
data \citep{Rasanen2015} and the reshift limitation of distance
indicators \citep{Zhou20}.

In the framework of the DSR, the purpose of this study is to assess
the constraints on the spatial curvature, which could be achieved by
confronting the currently largest standard candle quasar sample with
the largest compilation of SGL observations expected from the
forthcoming surveys. Specifically, the Large Synoptic Survey
Telescope (LSST) is expected to discover $\sim 10^5$
galaxy-scale lenses \citep{Oguri10,Vermai2019}, with the
corresponding source redshift reaching $z\sim 6$. In this paper, we
also take advantage of the recently compiled sample of quasar data
set comprising 1598 quasars covering the redshift range of
$0.036<z<5.100$ \citep{Risaliti2018}. Luminosity distances of
these new type of standard candles are inferred from the recent
method developed by \citet{Bisogni17}, based on the relation between
the UV and X-ray luminosities of high-redshift quasars. This paper
is organized as follows. In Sec. 2 and 3, we will briefly introduce
the methodology, strong gravitational lensing models, as well as the the
observational and simulated data in this analysis. In Sec. 4, we
show the forecasted constraints on the cosmic curvature. Finally, conclusions and discussions are summarized in Sec. 5.

\section{Methodology}
\subsection{Distance sum rule}

On the assumption of cosmological principle, one always turn to the
the FLRW metric to describe space-time of the Universe, which has the
following form (in units where $c=1$):
\begin{equation}\label{eq1}
ds^2=-dt^2+a^2(t)(\frac{1}{1-kr^2}dr^2+r^2d\Omega^2).
\end{equation}
Here $k$ is a constant ($k=+1$, $-$1, and 0 correspond to closed,
open, and flat universe) associated with the curvature parameter as
$\Omega_k=-k/a_0^2 H_0^2$, where $H_0$ denotes the Hubble constant.
Let us introduce dimensionless comoving distances $d_l\equiv d(0,
z_l)$, $d_s\equiv d(0, z_s)$ and $d_{ls}\equiv d(z_l, z_s)$.
For a galactic-scale strong lensing system, the
dimensionless comoving distance $d(z)$ between the lensing galaxy
(at redshift $z_l$) and the background source (at redshift $z_s$) is
given by
\begin{eqnarray}
\nonumber d(z_l, z_s)&=& (1+z_s)H_0 D_A(z_l,z_s)\\
&=&\frac{1}{\sqrt{|\Omega_k|}}f\left(
\sqrt{|\Omega_k|}\int^{z_s}_{z_l}\frac{H_0dz'}{H(z')} \right),
\end{eqnarray}
where
\begin{equation}\label{eq3}
f(x)=\left\{
   \begin{array}{lll}
   \sin(x)\qquad \,\ \Omega_k&<0, \\
   x\qquad\qquad \,\ \Omega_k&=0, \\
   \sinh(x)\qquad \Omega_k&>0. \\
   \end{array}
   \right.
\end{equation}
 In the
framework of FLRW metric, these distances are related via the
distance sum rule \citep{Bernstein06,Clarkson2008}
\begin{equation}\label{eq4}
d_{ls}={d_s}\sqrt{1+\Omega_kd_l^2}-{d_l}\sqrt{1+\Omega_kd_s^2}.
\end{equation}
Note that in terms of dimensionless comoving distances DSR will reduce to an
additivity relation $d_s =d_l+d_{ls}$ in the flat universe
($\Omega_k=0$). The source/lens distance ratios
$d_{ls}/d_s=D^{A}_{ls}/D^{A}_s$ can be assessed from the
observations of multiple images in SGL systems \citep{Cao2015b}.
Meanwhile, if the two other two dimensionless distances $d_l$ and
$d_s$ can be obtained from observations, the measurement of
$\Omega_k$ could be directly obtained \citep{Rasanen2015}. In this paper we will use the distance ratios $d_{ls}/d_s$ from the
simulated SGL sample representative of the data obtainable from the
forthcoming LSST survey \citep{Collett15}, while the distances on
cosmological scales ($d_l$ and $d_s$) will be inferred from the
recent multiple measurements of 1598 quasars calibrated as standard
candles \citep{Risaliti2018}.

\subsection{Strong gravitational lensing - distance ratio}

With the increasing number of detected SGL systems, strong
gravitational lensing has become an important astrophysical tool to
derive cosmological information from individual lensing galaxies,
with both high-resolution imaging and spectroscopic observations. In
this paper, we will focus on a method that can be traced back to
\citet{Futamase01} and furthermore extended in recent analysis
\citep{Bolton08,Cao2012,Cao2015b,Chen19} based on different SGL
samples. Specially, by combining the observations of SGL and stellar
dynamics in elliptical galaxies, one could naturally measure the
distance ratio $d_{ls}/d_s$, based on the measurements of Einstein
radius ($\theta_E$) and the central velocity dispersion
($\sigma_{lens}$) of the lens galaxies. The efficiency of such
methodology lies in its ability to put constraints on the
dynamic properties of dark energy \citep{Li16,Liu19}, the speed of
light at cosmological scales \citep{Cao18,Cao20}, and the validity
of the General Relativity at galactic scale
\citep{Cao17c,Collett18}. However, it should be stressed that
cosmological application of SGL requires a better knowledge of the
density profiles of early-type galaxies
\citep{Cao2016,Holanda17,Collett18}, the quantitative effect
which has been assessed with updated galactic-scale strong lensing
sample \citep{Chen19}. Therefore, to describe the structure
of the lens we will consider three types of models, which has
been extensively investigated in the literature
\citep{Qi2019,Zhou20}.

(I) Singular Isothermal Sphere (SIS) model: For the simplest SIS
model, the distance ratio is given by \citep{Koopmans06}.
\begin{equation} \label{SIE_E}
\frac{d_{ls}}{d_s}=\frac{c^2\theta_E}{4\pi \sigma_{SIS}^2}
=\frac{c^2\theta_E}{4\pi \sigma_{0}^2f_E^2},
\end{equation}
where $\sigma_{SIS}$ is the dispersion velocity due to SIS lens mass
distribution and $c$ the speed of light. In this analysis we also
introduce a free parameter $f_E$ to quantify the difference between
the SIS velocity dispersion ($\sigma_{SIS}$) and the observed
velocity dispersion of stars ($\sigma_{0}$), as well as other
possible systematic effects (see \citet{Ofek2003,Cao2012} for more
details).

(II) Power-law model: Motivated by recent studies supporting
non-negligible deviation from SIS for the slopes of density profiles
of individual galaxies \citep{Koopmans06,Humphrey10,Sonnenfeld13a},
we choose to generalize the SIS model to a spherically symmetric
power-law mass distribution ($\rho\sim r^{-\gamma}$). So the
distance ratio for a power-law lens model can be written as
\citep{Ruff2011,Koopmans06,Bolton2012}
\begin{equation} \label{sigma_gamma}
\frac{d_{ls}}{d_s}=\frac{c^2\theta_E}{4\pi
\sigma_{ap}^2}\left(\frac{\theta_{ap}}{\theta_E}\right)^{2-\gamma}f^{-1}(\gamma),
\end{equation}
where $f(\gamma)$ is a certain function of the radial mass profile slope (see e.g. \citep{Cao2015b} for details),
while the luminosity averaged line-of-sight velocity dispersion
$\sigma_{ap}$ can be measured inside the circular aperture of the
angular radius $\theta_{ap}$. Note that SIS lens
model corresponds to $\alpha=2$.

(III) Extended power-law model: Considering the possible difference
between the luminosity density profile ($\nu(r)\sim r^{-\delta}$)
and the total-mass (i.e. luminous plus dark-matter) density profile
($\rho(r)\sim r^{-\alpha}$), one may solve the radial Jeans equation
in spherical coordinate system to derive the dynamical mass inside
the aperture radius \citep{Koopmans05}. Therefore, the distance
ratio -- in the framework of this complicated lens profile -- will
be straightforwardly obtained, through the combination of dynamical
mass and lens mass within the Einstein radius \citep{Chen19}
\begin{eqnarray}\label{sigma_alpha_delta}
\nonumber
\frac{d_{\rm ls}}{d_{\rm s}}&=& \left(\frac{c^2}{4\sigma_{ap}^2}\theta_{\rm E}\right)\frac{2(3-\delta)}{\sqrt{\pi}(\xi-2 \beta)(3-\xi)} \left( \frac{\theta_{\rm ap}}{\theta_{\rm E}}\right)^{2-\alpha}\\
&\times&\left[\frac{\lambda(\xi)-\beta\lambda(\xi+2)}{\lambda(\alpha)\lambda(\delta)}\right]~,
\end{eqnarray}
where $\xi=\alpha+\delta-2$,
$\lambda(x)=\Gamma(\frac{x-1}{2})/\Gamma(\frac{x}{2})$. Note that
$\delta=\alpha$ denotes that the shape of the luminosity density
follows that of the total mass density, i.e., the power-law lens
model. Moreover, in this model, a new parameter $\beta(r) = 1 -
{\sigma^2_t} / {\sigma^2_r}$ is included to quantify the anisotropy of stellar velocity.  We assume that it follows a Gaussian distribution of
$\beta=0.18\pm0.13$ suggested by recent observations of several
nearby early-type galaxies \citep{Gerhard01,Schwab2009}.

\subsection{Distance calibration from high-redshift quasars}

In the past decades, great efforts have been made in investigating
the ``redshift - luminosity distance" relation in quasars for the
purpose of cosmological studies, based on different relations
involving the quasar luminosity \citep{Baldwin77,Watson11,Wang13}.
In particular, the non-linear relation between the X-ray and UV
luminosities of quasars looked very promising \citep{Avni1986}.
However, suffering from the extreme variability and a wide range of
luminosity, it still remains controversial whether quasars can be
classified as "true" standard (or standardizable) candles in the
Universe. Meanwhile, it should be pointed out that high scatter in
the observed relations or the limitation of poor statistics remain
the major uncertainties in most of these methods. Attempting to use
these quasars by virtue of the non-linear relation between the X-ray
and UV luminosities, one is usually faced with the challenge of
large dispersions and observational biases. A key step forward was
recently made by \citet{Risaliti2018}, who gradually refined the
selection technique and flux measurements, which provided a suitable
subsample of quasars (with an intrinsic dispersion smaller than 0.15
dex) to measure the luminosity distance.

Following the approach described in \citet{Risaliti2015}, there exits a
relation between the luminosities in the X-rays ($L_X$) and UV band
($L_{UV}$)
\begin{equation}
\log(L_X)=\hat{\gamma}\log(L_{UV})+\beta',
\end{equation}
where $\hat{\gamma}$ and $\beta'$ denote the slope parameter and the
intercept. Combing Eq.~(8) with the well-known expression of
$L=F\times 4\pi D_L^2$, the luminosity distance can be rewritten as
\begin{equation}
\log(D_L)=\frac{1}{2-2\hat{\gamma}}\times[\hat{\gamma}\log(F_{UV}) - \log(F_X) + \hat{\beta}],
\end{equation}
a function of the respective fluxes ($F$), the slope parameter
($\hat{\gamma}$) and the normalization constant
($\hat{\beta}=\beta'+(\hat{\gamma}-1)\log_{10} 4\pi$). Therefore,
from theoretical point of view, the luminosity distance can be
directly determined from the measurements of the fluxes of $F_X$ and
$F_{UV}$, with a reliable knowledge of the dispersion $\delta$ in
this relation and the value for the two parameters ($\hat{\gamma}$,
$\hat{\beta}$) characterizing the $L_X - L_{UV}$ relation. However,
it has been established that the $L_X - L_{UV}$ relation was
characterized by a high dispersion. Through the analysis of
different quasar samples with multiple observations available,
previous works derived a consistent value for the slope parameter
($\hat{\gamma}=0.599\pm0.027$) and the intrinsic dispersion of the
relation ($0.35\sim0.40$ dex) \citep{Lusso2010,Young2010}. It was
found in subsequent analysis quantifying the observational effects
\citep{Lusso2016} that the magnitude of the intrinsic dispersion can
be eventually decreased to the level of $<0.15$ dex. They identified
a subsample of quasars without the major contributions from
uncertainties in the measurement of the (2keV) X-ray flux,
absorption in the spectrum in the UV and in the X-ray wavelength
ranges, variability of the source and non-simultaneity of the
observation in the UV and X-ray bands, inclination effects affecting
the intrinsic emission of the accretion disc, and the selection
effects due to the Eddington bias \citep{Risaliti2018}. Besides the
lower dispersion in the relation, the reliability and effectiveness
of the method strongly depend on the lack of evolution of the
relation with redshift \citep{Bisogni17}. Finally,
\citet{Risaliti2018} produced a final, high-quality catalog of 1598
quasars, by applying several filters (X-ray absorption,
dust-reddening effects, observational contaminants in the UV,
Eddington bias) to the parent sample from the Sloan Digital Sky
Survey (SDSS) quasar catalogues \citep{Shen2011,Paris2017} and the
XMM-Newton Serendipitous Source Catalogue \citep{Rosen2016}. The
final results indicated that such refined selection of the sources
could effectively mitigate the large dispersion in the $L_X -
L_{UV}$ relation, with a tractable amount of scatter avoiding
possible contaminants and unknown systematics (see
\citet{Risaliti2018} for more details). More importantly, the
similar analysis has supported the non-evolution of $L_X - L_{UV}$
relation with the redshift, which is supported by the subsequent
study involving the intercept parameter $\hat{\beta} = 8.24\pm0.01$,
the slope parameter $\hat{\gamma}=0.633\pm0.002$, and smaller
dispersion $\hat{\delta}=0.24$ in a new, larger quasar sample
\citep{Risaliti2018}.

Therefore, with the gradually refined selection technique and flux
measurements, as well as the elimination of systematic errors caused
by various aspects, their discovery has a major implication: based
on a Hubble diagram of quasars, new measurements of the expansion
rate of the Universe could be obtained in the range of
$0.036<z<5.10$.

\begin{figure}
\begin{center}
\includegraphics[width=0.9\linewidth]{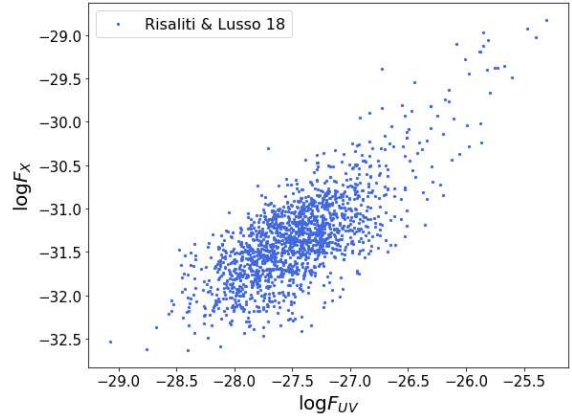}
\end{center}
\caption{Scatter plot of the flux measurements of 1598 quasars
\citep{Risaliti2018}.}
\end{figure}

\begin{figure*}
\begin{center}
\includegraphics[width=0.4\linewidth]{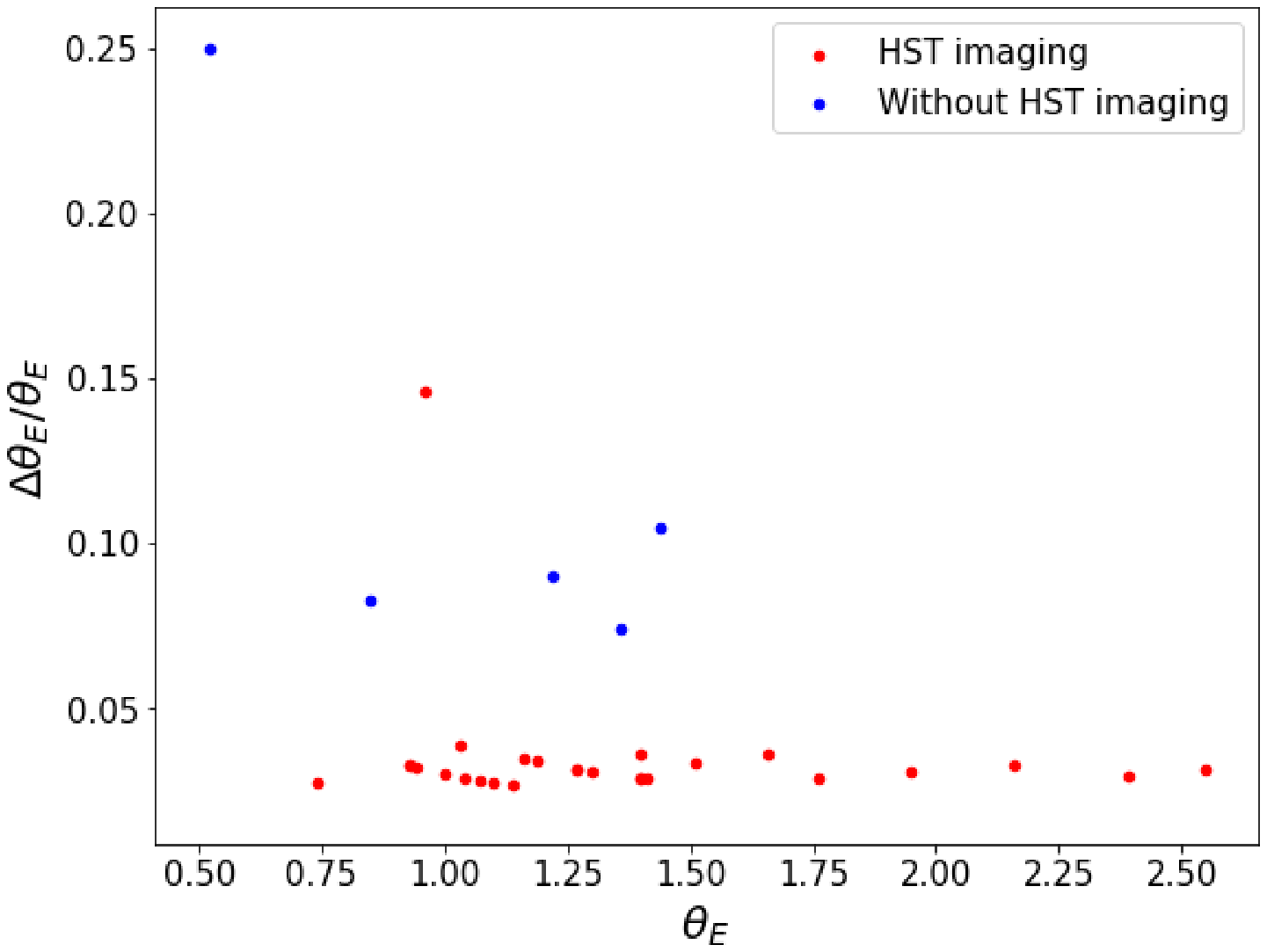}\includegraphics[width=0.42\linewidth]{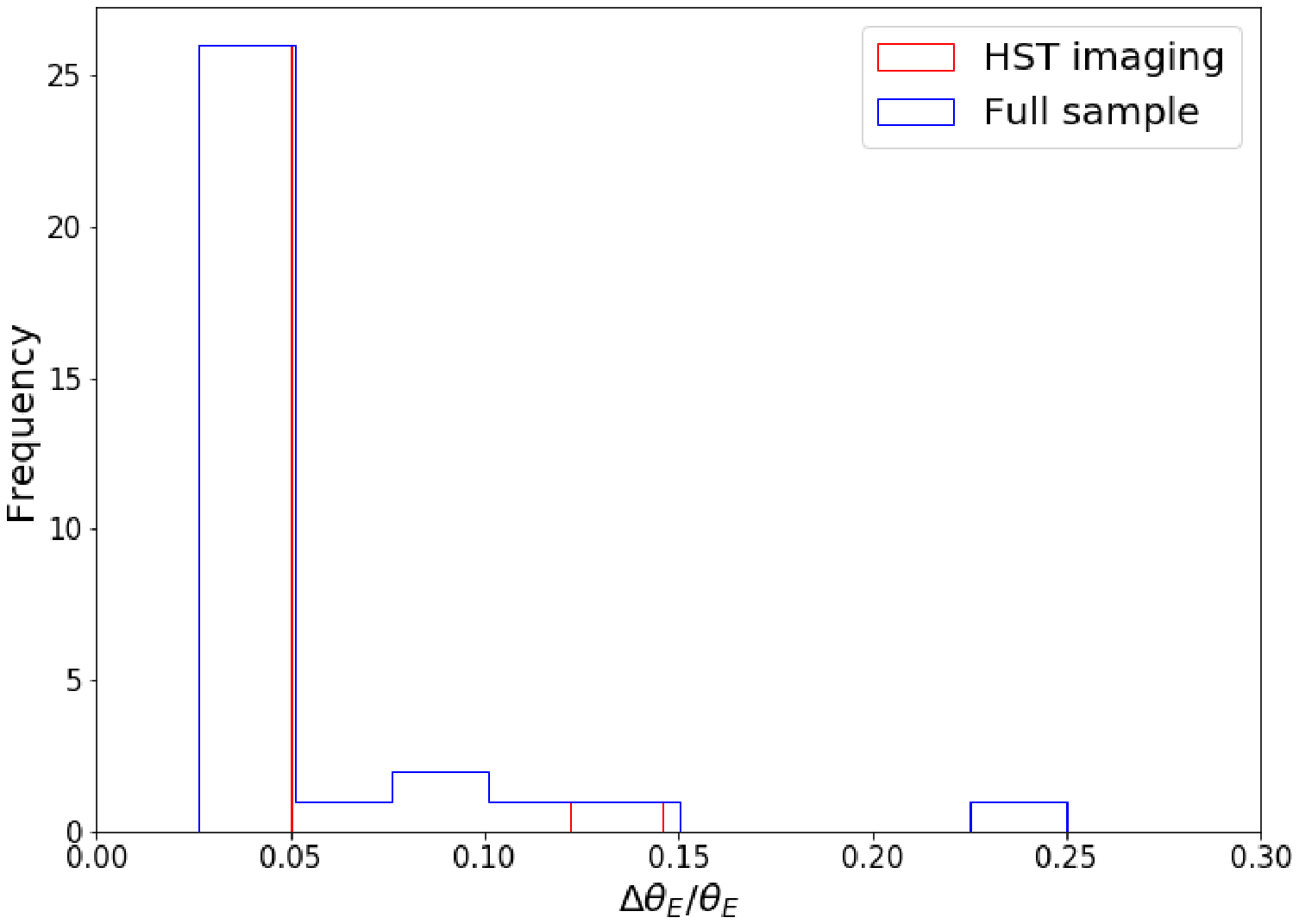}
\end{center}
\caption{\textbf{Fractional uncertainty of the Einstein radius
($\Delta\theta_E/\theta_E$) determination as a function of the
Einstein radius ($\theta_E$)(left panel) and the corresponding
histogram plot (right panel), based on the SL2S sample with HST
imaging and HST+CFHT imaging.}}
\end{figure*}

\section{Observations and simulations}

\subsection{The observational quasar data}

In this paper, we turn the improved ``clean" sample including 1598
quasars, with reliable measurements of intrinsic X-ray and UV
emissions assembled in \citet{Risaliti2018}. The flux measurements
concerning X-ray and UV emissions with the final sample is shown in
Fig.~1. Possible cosmological application of these standard
candles has recently been discussed in the literature
\citep{Melia19}. More recently, the multiple measurements of
high-redshift quasars have been used for testing the cosmic
distance duality relation (CDDR), based on the relation between the
UV and X-ray luminosities of quasars, combined with the VLBI
observations of compact structure in radio quasars \citep{Zheng20}.

According to the Eq.~(6), we would be able to derive luminosity
distances $D_L(z)$ and hence dimensionless co-moving distances of
the lens $d_l$ and the source $d_s$ for each SGL system, from UV and
X-ray fluxes of the quasars whose redshifts are equal to $z_l$ and
$z_s$, respectively. However, there are two potential problems to be
solved. First is a high intrinsic scatter ($\hat{\delta}$) in the
quasars sample, based on the UV and X-ray flux measurements. Second
is that $\hat{\beta}$ and $\hat{\gamma}$ parameters are unknown.
Fortunately, \citet{Melia19} used the quasars sample to achieve
cosmological test without any external calibrator, treating the
slope $\hat{\gamma}$, the intercept $\hat{\beta}$, and the intrinsic
scatter $\hat{\delta}$ as free parameters to be fit. It was revealed
that the quasar data can be self-calibrated under such individual
optimization within a specified cosmology. For example in
$\Lambda$CDM model one obtains: $\hat{\gamma}=0.639\pm 0.005$,
$\hat{\beta}=7.02\pm 0.012$, $\hat{\delta}=0.231\pm0.0004$, and
$\Omega_m=0.31\pm 0.05$. In this analysis, we will not confine
ourselves to any specific cosmology, but instead we reconstruct the
dimensionless co-moving distance function $d(z)$, modeled as a
polynomial expansion in $z$ or logarithmic polynomial expansion of
$\log(1+z)$ \citep{Rasanen2015,Liao2017,Li2018}. For the first case,
the dimensionless angular diameter distance is parameterized by a
third-order polynomial function of redshift
\begin{equation}
d(z)=z+a_1z^2+a_2z^3,
\end{equation}
with the initial conditions of $d(0)=0$ and $d^{'}(0)=1$. For the
second case, we perform empirical fit to the quasar measurements,
based on a third-order logarithmic polynomial of
\begin{equation}
d(z)=ln(10)(x+b_1x^2+b_2x^3),
\end{equation}
with $x=\log(1+z)$. Note that in the above two parameterizations,
($a_1$, $a_2$) and ($b_1$, $b_2$) represent two sets of constant
parameters that need to be optimized and determined by flux
measurements data in X-ray and UV emissions. Meanwhile, the
logarithmic parametrization, benefiting from a more rapid
convergence at high redshifts with respect to the standard linear
parametrization, has proved to be a more reasonable approximation at
high redshifts \citep{Risaliti2018}.

\begin{figure*}
\begin{center}
\includegraphics[width=0.42\linewidth]{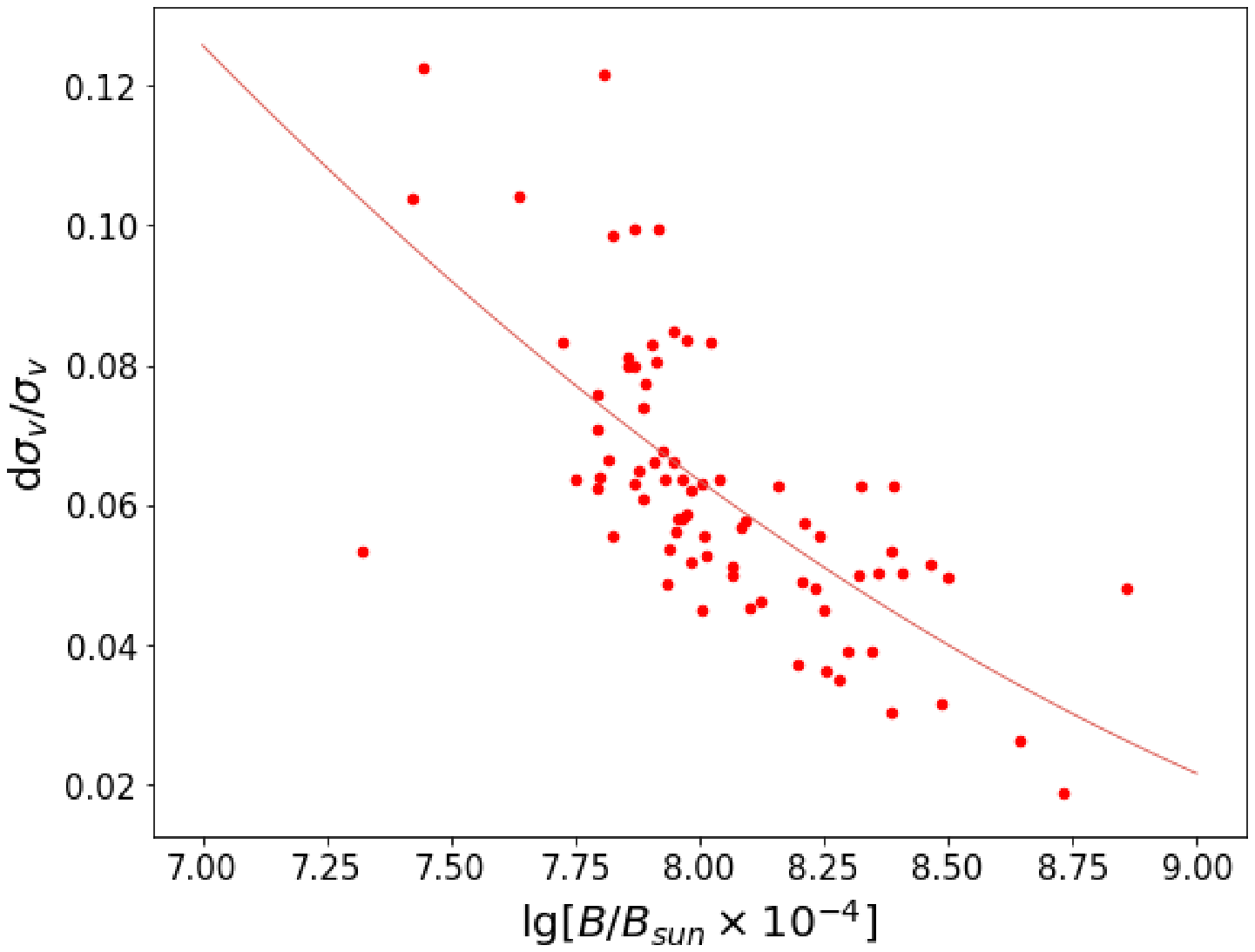}\includegraphics[width=0.4\linewidth]{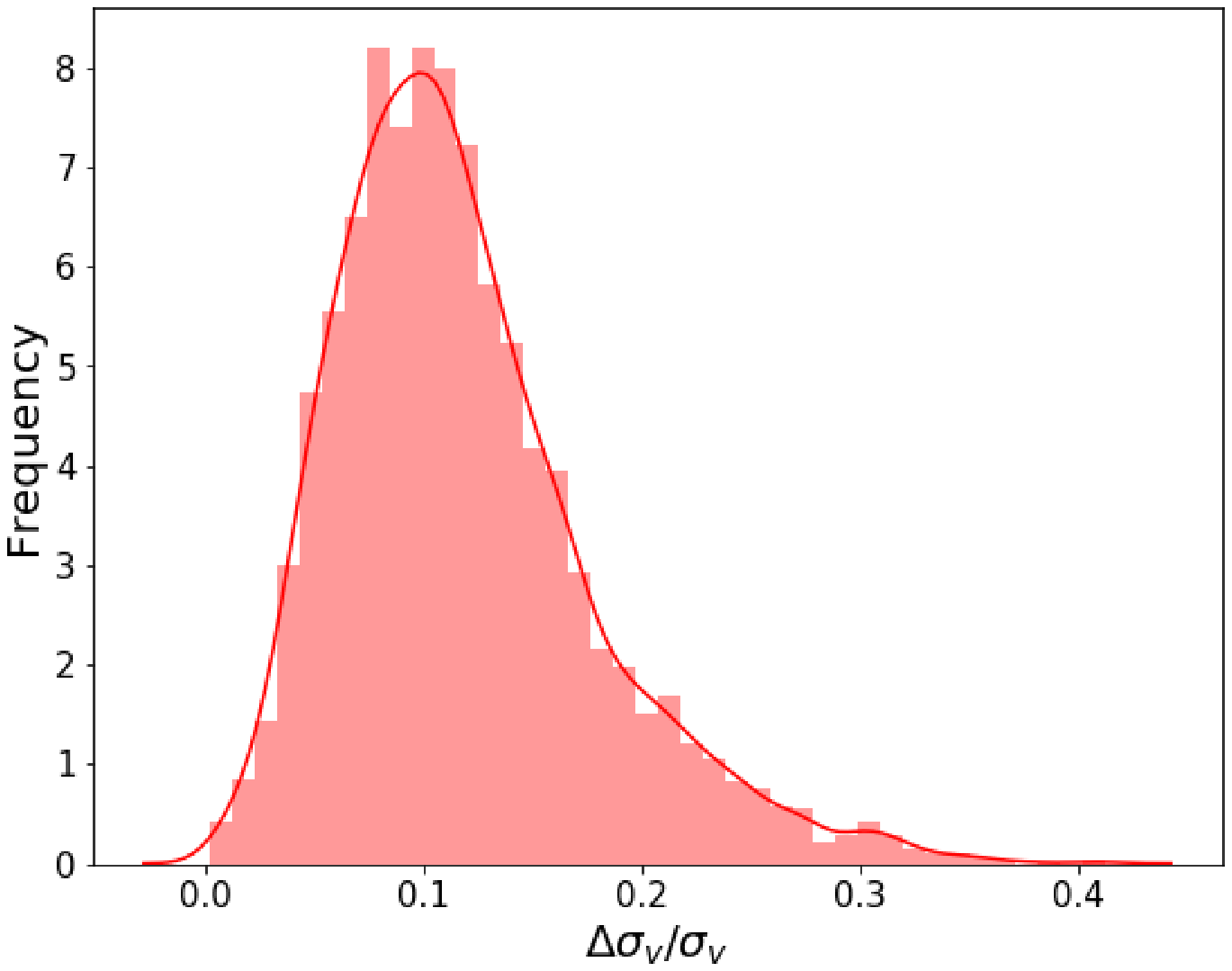}
\end{center}
\caption{Left panel: Fractional uncertainty of the velocity
dispersion ($\Delta\sigma_v/\sigma_v$) as a function of the lens
surface brightness ($B$) for the SLACS sample, with the best-fitted
correlation function denoted as the red solid line. Right panel: The
distribution of the velocity dispersion uncertainty for the
simulated SGL sample.}
\end{figure*}

\begin{figure*}
\begin{center}
\includegraphics[width=0.43\linewidth]{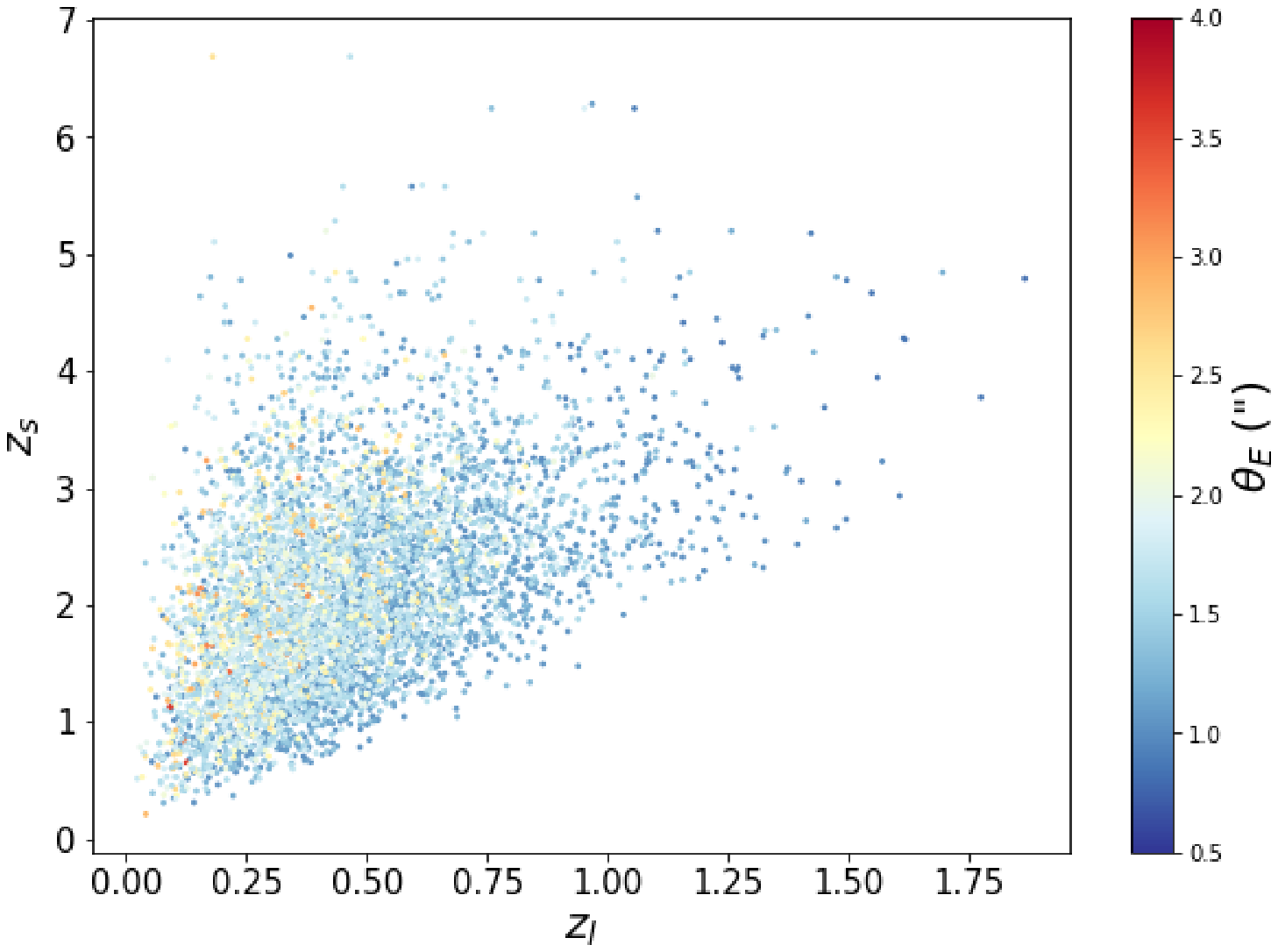}\includegraphics[width=0.43\linewidth]{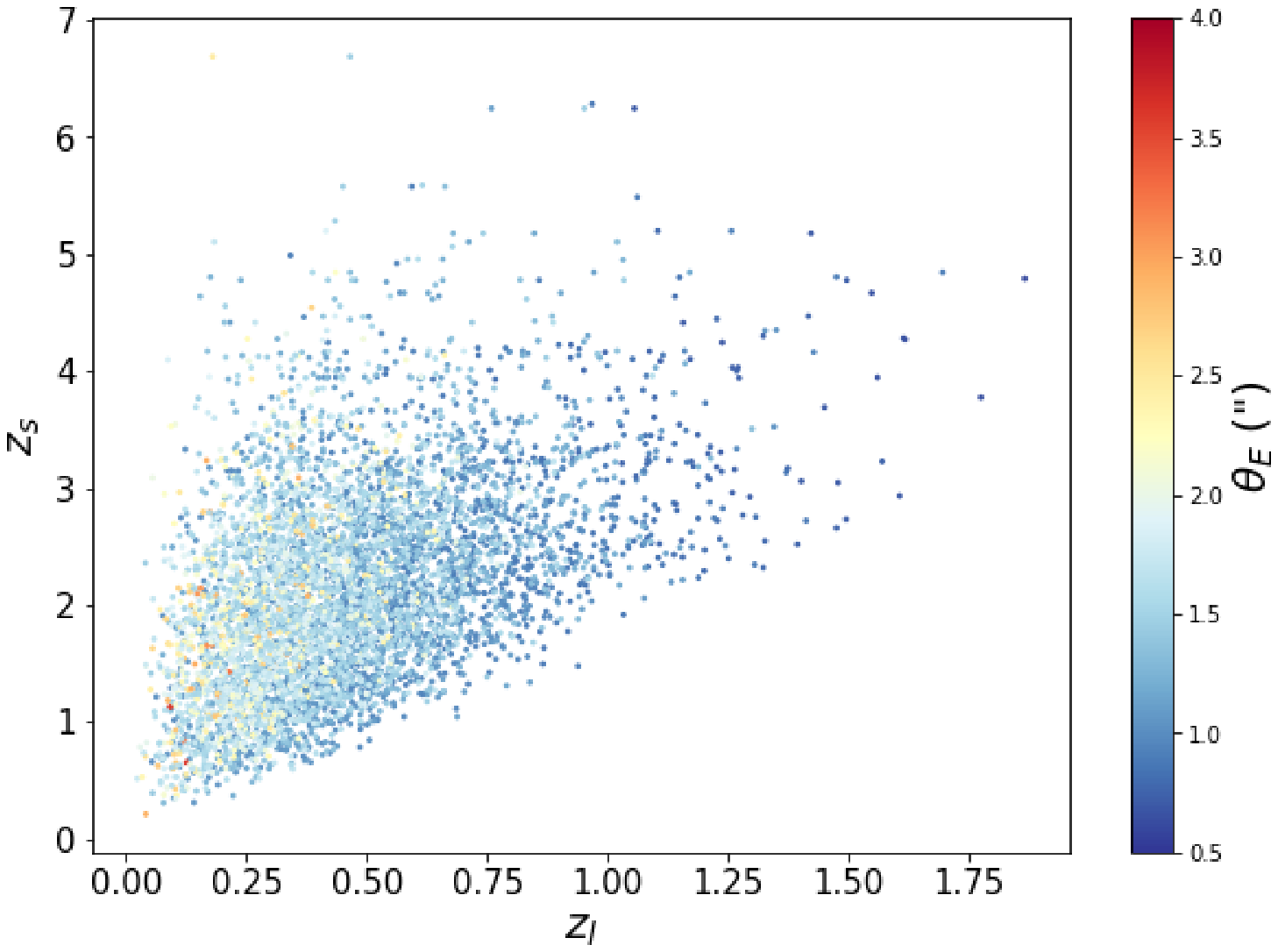}
\end{center}
\caption{\textbf{The scatter plot of the simulated lensing systems
based on the standard polynomial model (left panel) and logarithmic
polynomial model (right panel), with the gradient color denoting the
value of the Einstein radius.} }
\end{figure*}

In order to reconstruct the profile of dimensionless co-moving
distance $d(z)$ function up to the redshifts $z=5.0$, we make use of
the publicly available code called the \emph{emcee}
\footnote{https://pypi.python.org/pypi/emcee} \citep{Foreman 2013},
to obtain the best-fit values and the corresponding $1\sigma$
uncertainties of relevant parameters ($a_1$, $a_2$, $b_1$ and $b_2$
in our case). It is worth stressing here that one may worry that the
cosmographic expansions are only valid at low redshift. However, the
recent analysis of high-redshift Hubble diagram indicated that these
relations are valid beyond $z\sim 4$, although fitting a log
polynomial cosmography may hide certain features of the quasar data
\citep{Yang19}. Meanwhile, our results demonstrate that a
third-order polynomial function adopted in \citet{Risaliti2018} is
sufficient enough to expand the luminosity distance, since the
inclusion of higher orders in the polynomial expansion have
negligible effect on the final reconstruction results. The
chi-square $\chi^2$ objective function we minimized is defined as
\begin{equation}
\chi^2 = \sum_{i=1}^{1598} \frac{[\log(F_{X,i})-\Psi_{th} ([F_{UV}]_i;D_L[z_i])]^2}{\sigma_{F_{X,i}}^2+\hat{\delta}^2},
\end{equation}
where $\hat{\delta}$ represents the global intrinsic dispersion, the
$\sigma_{F_{X,i}}$ denotes the i-th measurement error of flux
$F_{X,i}$ in X-ray waveband. The function $\Psi_{th}$ is defined as
\begin{equation}
\Psi_{th}=\hat{\beta}+\hat{\gamma}
\log(F_{UV,i})+2(\hat{\gamma}-1)\log(D_L(z_i)),
\end{equation}
in terms of the measured fluxes ($F_{X,i}$, $F_{UV,i}$) and the
luminosity distance $D_L(z)=c/H_0(1+z)d(z)$. It should be pointed
out that the measurement error of the flux in UV band is ignored in
this analysis since $\sigma_{F_{UV,i}}$ is insignificant comparing
with $\sigma_{F_{X,i}}$ and $\hat{\delta}$. Meanwhile, we have also
assumed a fiducial value for the Hubble constant $H_0=67.4$ km
s$^{-1}$ Mpc$^{-1}$, based on the results obtained from Planck 2018
data (TT, TE, EE+lowE+lensing) \citep{Planck Collaboration}. For the
first case, the best-fit quasar parameters and the 68\% C.L. are
determined as $\hat{\gamma}=0.613\pm0.011$,
$\hat{\beta}=7.970\pm0.312$, and $\hat{\delta}=0.230\pm0.003$. The
corresponding results will change to $\hat{\gamma}=0.616\pm0.011$,
$\hat{\beta}=7.530\pm0.283$, and $\hat{\delta}=0.230\pm0.003$ for the
second case.

\subsection{The simulated SGL data from LSST}

It is broadly reckoned that the future wide-area and deep surveys,
such as the Large Synoptic Survey Telescope \citep{Vermai2019} and
the Dark Energy Survey (DES) \citep{Frieman2004} will revolutionize
the strong lensing science, by
increasing the number of known galactic lenses by orders of
magnitude. More specifically, the forthcoming photometric LSST
survey will discover $\sim 10^5$ strong gravitational lenses
\citep{Collett15}, the cosmological application of which has become
the focus of the forecasted yields of LSST in the near future
\citep{Cao17c,Cao18,Ma19,Cao20}.

Based on the publicly available simulation programs
\footnote{github.com/tcollett/LensPop} explicitly described in
\citet{Collett15}, we simulate a realistic population of strong
lensing systems with early-type galaxies acting as lenses. The
singular isothermal sphere (SIS) is adopted to model the mass
distributions of the lensing galaxies, the number density of which
is characterized by the velocity dispersion function (VDF) from the
measurements of SDSS Data Release 5 \citep{Choi07}. Now one
important issue should be emphasized in our simulations. In order to
achieve our $\Omega_k$ test with the combination of strong lensing
and stellar dynamics, valuable additional information such as
spectroscopic redshifts ($z_l$ and $z_s$) and spectroscopic velocity
dispersion ($\sigma_{ap}$) are necessary. Since these dedicated
observations and substantial follow-up efforts for a sample of
$10^5$ SGL systems are expensive, it is more realistic to focus only
on a particular well-selected subset of LSST lenses, as was proposed
in the recent discussion of multi-object and single-object
spectroscopy to enhance Dark Energy Science from LSST
\citep{Hlozek19,Mandelbaum19}. Therefore, in our analysis the final
SGL sample is restricted to 5000 elliptical galaxies with the
velocity dispersion of 200 km/s $<\sigma_{ap} <$ 300 km/s, following
the recent investigation of medium-mass lenses to minimize the
possible discrepancy between Einstein mass and dynamical mass for
the SIS model \citep{Cao2016}. The final simulated results show that
the distributions of velocity dispersions and Einstein radii are
very similar to those of the current SL2S sample
\citep{Sonnenfeld13}.

\begin{figure}
\begin{center}
\includegraphics[width=0.95\linewidth]{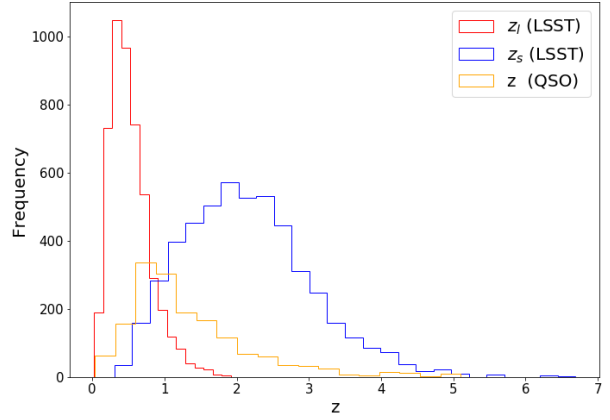}
\end{center}
\caption{Redshift distribution of quasars used to assess distances
and SGL systems from future LSST survey.}
\end{figure}

Concerning the uncertainty budget, LSST could provide high-quality
(sub-arcsecond) imaging data in general, especially in the $g$-band.
However, in order to extract the full potential of LSST, obtaining
high-resolution images for the lensing systems could also require
additional imaging data from space-based facilities (HST), with
detailed follow-up of individual SGL systems. Meanwhile, the
participation of other ground-based facilities makes it possible to
derive additional spectroscopic information, i.e., lens redshifts,
source redshifts, and velocity dispersion measurements for
individual lenses. In this analysis, different strategies will be
applied to cope with the fractional uncertainty of the Einstein
radius and stellar velocity dispersion, considering the possible
correlations between the observational precision and other intrinsic
properties of the lensing system (such as the mass or the brightness
of the lens).

For the uncertainty of the Einstein radius, we turn to 32
SGL systems recently detected by Strong Lensing Legacy Survey
(SL2S), with Canada¨CFrance¨CHawaii Telescope (CFHT) near-infrared
ground-based images or Hubble Space Telescope (HST) imaging data
\citep{Sonnenfeld13}. The HST imaging data were taken with the
Advanced Camera for Surveys (ACS; filters: F814W/F606W; exposure
time: 800/400s), Wide Field and Planetary Camera 2 (WFPC2; filter:
F606W; exposure time: 1200s), and Wide Field Camera 3 (WFC3;
filters: F600LP/F475X; exposure time: 720s), which have been
observed with HST as part of programs 10876, 11289 (PI: J. P. Kneib)
and 11588 (PI: R. Gavazzi). In addition to space-based photometry,
the NIR images for some of the SL2S lenses were observed with the
instrument WIRCam \citep{Puget04} in the $K_s$, $J$ and $H$ bands.
We refer the reader to \citet{Sonnenfeld13} for more detailed
information of the CFHT observations for each target (exposure time,
etc.). The scatter and histogram plots of the fractional uncertainty
of Einstein radius are respectively shown in Fig.~2, concerning the
full sample with HST+CFHT imaging and the sub-sample with HST
imaging. Not surprisingly, most of the lenses with high-precision
Einstein radius measurements are derived from systems with HST
data. Focusing on the full catalogue of SGL systems, one can
clearly see a possible correlation between the fractional
uncertainty of the Einstein radius and $\theta_E$, i.e. the lenses
with smaller Einstein radii would suffer from large $\theta_E$
uncertainty, as reported previously in the previous strong lensing
analysis. Meanwhile, for the full sample with HST+CFHT
imaging data (i.e., CFHT image when HST image is not available), the
fractional uncertainty of the Einstein radius is taken at the level
of 8\%, 5\% and 3\% (the mean uncertainty within each certain
$\theta_E$ bin) for small Einstein radii lenses
($0.5"<\theta_E<1.0"$), intermediate Einstein radii lenses
($1"\leq\theta_E<1.5"$), and large Einstein radii lenses
($\theta_E\geq1.5$). Such error strategy will be implemented in the
simulations of our LSST lens sample. Meanwhile, in the optimistic
case, i.e. when all of the LSST lenses considered in this work will
be observed with HST-like image quality, it is reasonable to take
the fractional uncertainty of the Einstein radius at a level of 3\%
\citep{Hilbert09}. For the uncertainty of the velocity dispersion,
we turn to 70 SGL systems observed in the Sloan Lens ACS survey
(SLACS) \citep{Bolton08} and quantitatively analyze its correlation
with the lens surface brightness in the $i$-band. The
population of strong lenses is dominated by galaxies with velocity
dispersion of $\sigma_{ap}\sim 230$ km/s (median value), while the
Einstein radius distribution is characterized by the median value of
$\theta_E=1.10"$. Such restricted SLACS lens sample, which falls
within the velocity dispersion criterion applied in this analysis
(200 km/s $<\sigma_{ap} <$ 300 km/s), is a good representative
sample of what the future LSST survey might yield. As can be
clearly seen from the results shown in Fig.~3, strong evidence of
anti-correlation between these two quantities is revealed in our
analysis. Using the best-fitted correlation function derived from
the current SGL sample, we obtain in Fig.~3 the distribution of
velocity dispersion uncertainty for the lenses discoverable in
forthcoming LSST survey, which is well consistent with the previous
strategy of assigning an overall error of 5\% on $\sigma_{ap}$
\citep{Cao2015b,Zhou20}.

It should be pointed out that LSST will discover a number of
fainter, smaller-separation lenses where it is not clear that the
same level of precision can be reached. Therefore, two selection
criteria of the Einstein radius and the $i$-band magnitude are
applied to our particular well-selected subset of \textbf{LSST
lenses ($\theta_E>0.5"$ and $m_i<22$).} In this paper, we generate
two SGL samples using the standard polynomial and logarithmic
polynomial cosmographic reconstructions (taking the best fitted
parameters of these reconstructions). The scatter plots of the
simulated lensing systems based on standard polynomial and
logarithmic polynomial cosmographic reconstructions are shown in
Fig.~4. For a good comparison, Fig.~5 illustrates the redshift
coverage of the current quasar sample and simulated SGL sample,
which demonstrates the perfect consistency between the redshift
range of high-$z$ quasars and LSST lensing systems.

\begin{table*}
\begin{center}
\begin{tabular}{c| c c c c c}
\hline
 Standard polynomial   & $\Omega_k$ & $f_E$ &$\gamma$ &$\alpha$ &$\delta$  \\
\hline
 SIS    & $0.002\pm0.035$ & $1.000\pm0.002$ & $\Box$ & $\Box$ & $\Box$ \\
\hline
 Power-law spherical  & $-0.007\pm0.029$ & $\Box$  &$2.000\pm0.012$ & $\Box$  & $\Box$ \\
\hline
Extended power-law   & $0.003\pm0.045$ & $\Box$  & $\Box$ &$2.000\pm0.014$ &$2.171\pm0.035$ \\
\hline
 Power-law spherical (with HST imaging) & $-0.008\pm0.028$ & $\Box$  &$2.000\pm0.012$ & $\Box$  & $\Box$ \\
\hline \hline
logarithmic polynomial        & $\Omega_k$ & $f_E$  &$\gamma$ &$\alpha$ &$\delta$  \\
\hline

 SIS    & $-0.001\pm0.030$ & $1.000\pm0.003$ & $\Box$ & $\Box$ & $\Box$ \\
\hline

 Power-law spherical  & $-0.007\pm0.016$ & $\Box$ &$2.000\pm0.013$ & $\Box$ & $\Box$\\
\hline

Extended power-law   & $0.002\pm0.031$ & $\Box$  & $\Box$ &$2.002\pm0.016$ &$2.172\pm0.035$ \\
\hline

\end{tabular}
\caption{Constraints on the cosmic curvature and lens profile
parameters for three types of lens models, in the framework of
standard polynomial and logarithmic polynomial cosmographic
reconstructions.} \label{SIE_table}
\end{center}
\end{table*}

\section{Results and discussion}

The constraints on the cosmic curvature, based on the simulated SGL
systems supplemented with the constructed Hubble diagram of
high-redshift quasars, are obtained by maximizing the likelihood
${\cal L} \sim \exp{(-\chi^2 / 2)}$. In our analysis, $\chi^2$ is
constructed as
\begin{equation}
\chi^2(\textbf{p},\Omega_k)=\sum_{i=1}^{N} \frac{\left({\cal
D}_{th}({z}_i;\Omega_k)- {\cal
D}_{obs}({z}_i;\textbf{p})\right)^2}{\sigma_{\cal D}(z_i)^2},
\end{equation}
where ${\cal D} = d_{ls}/d_s$ and $N$ is the number of the data
points. The theoretical distance ratio $\mathcal{D}_{th}$ dependent
on $\Omega_k$ is given by Eq.~(\ref{eq4}), while its observational
counterpart is dependent on the lens model adopted
Eq.~(\ref{SIE_E}), (\ref{sigma_gamma}) and
(\ref{sigma_alpha_delta}). Free parameters in these lens model are
collectively denoted as $\textbf{p}$, and $\sigma_D$ stands for the
uncertainty of the distance ratio expressed as
$\sigma_D^2=\sigma_{SGL}^{2}+\sigma_{QSO}^2$. Note that the
statistical error of SGL ($\sigma_{SGL}$) is propagated from the
measurement uncertainties of the Einstein radius and velocity
dispersion, while $\sigma_{QSO}$ depends on the uncertainties of
$d(z)$ function (polynomial and log-ploynomial parameterized
distance) reconstructed from the quasars. As previously mentioned,
the aim of this work is to estimate the cosmic curvature by
combining the constructed Hubble diagram of high-redshift quasars
with galactic-scale strong lensing systems expected to be seen by
the forthcoming LSST survey. Therefore, our analysis will be
performed on two different reconstruction schemes: the standard
polynomial cosmographic reconstruction and the logarithmic
polynomial reconstruction. The numerical results for the cosmic
curvature $\Omega_k$ and lens model parameters are summarized in
Table 1, with the marginalized distributions with 1$\sigma$ and
2$\sigma$ contours shown in Fig.~7.

\begin{figure*}
\begin{center}
\includegraphics[width=0.33\linewidth]{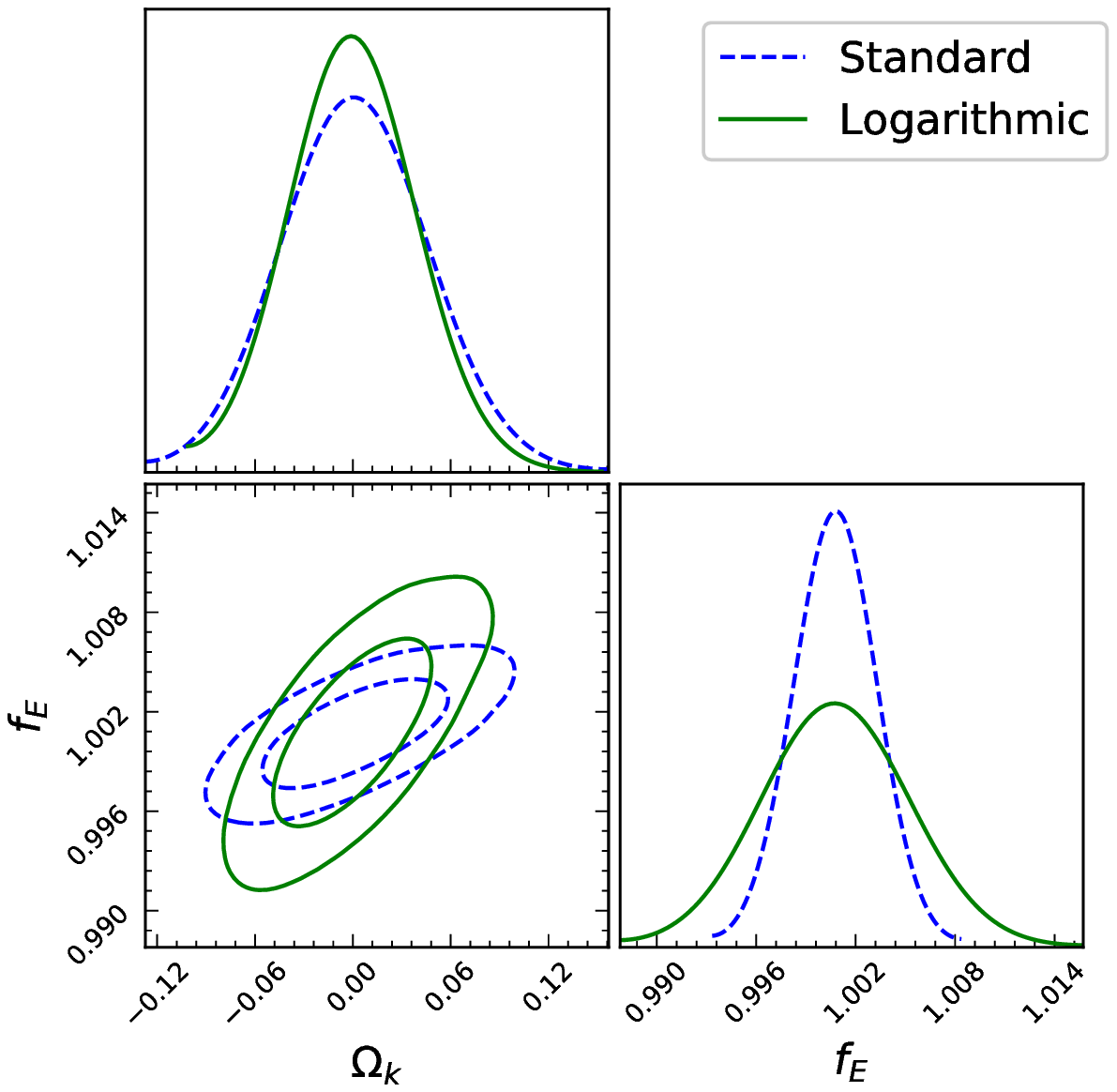} \includegraphics[width=0.33\linewidth]{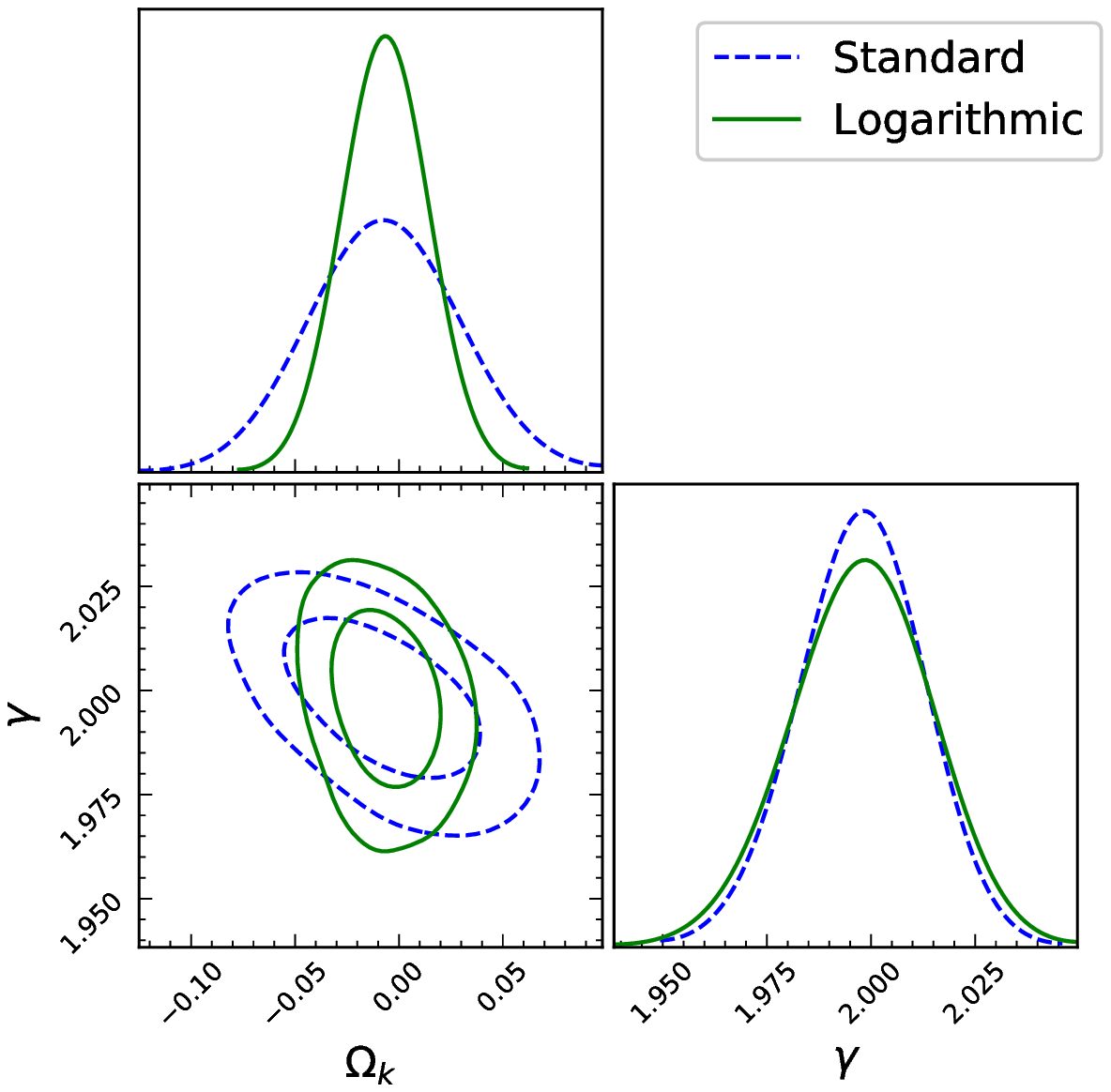}
\includegraphics[width=0.33\linewidth]{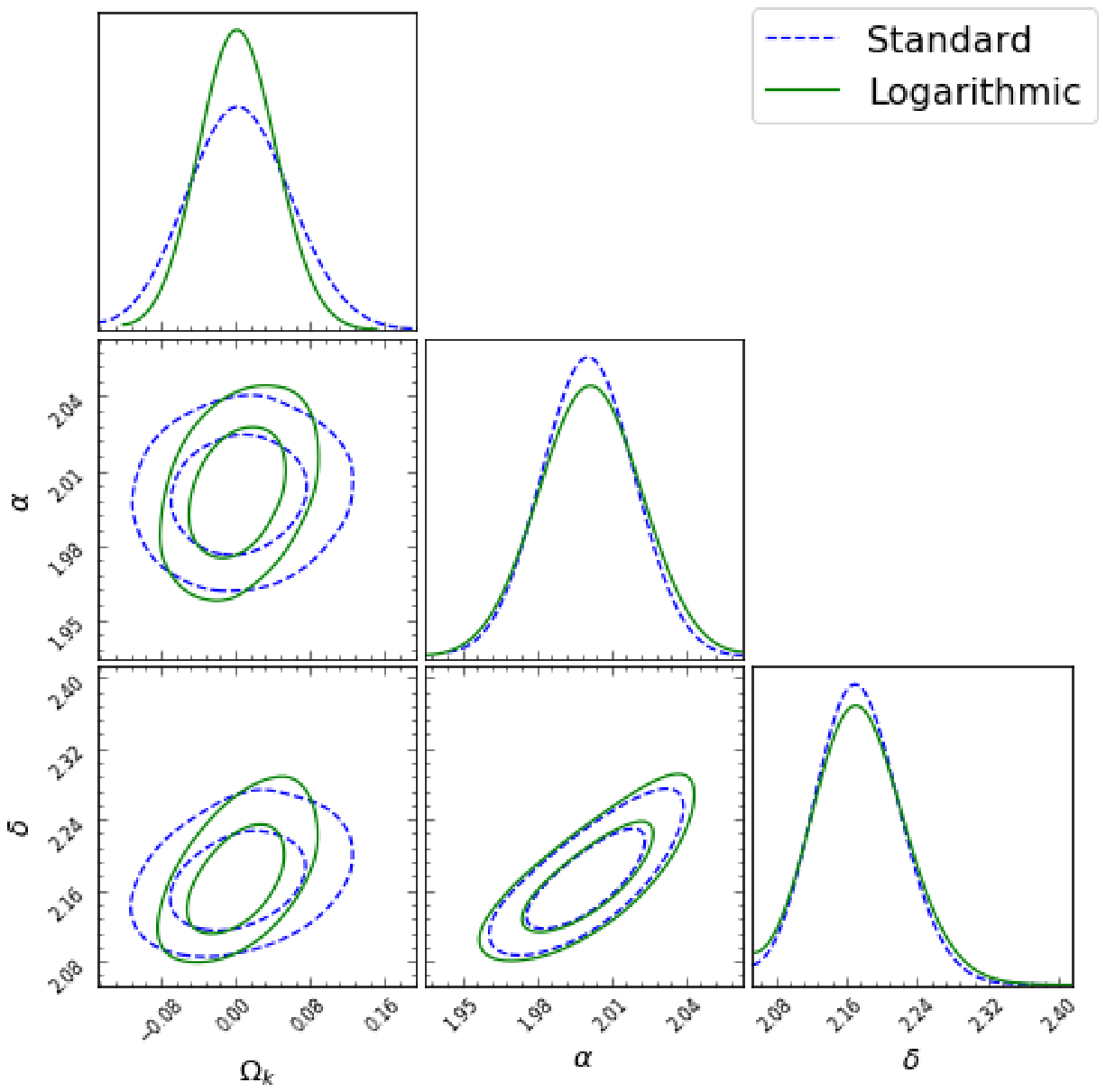}
\end{center}
\caption{The 2-D regions and 1-D marginalized distribution with the
1-$\sigma$ and 2-$\sigma$ contours of all parameters from the
standard polynomial (blue dotted line) and the logarithmic
polynomial (green solid line) cosmographic reconstruction, in the
framework of three lens models: SIS (left), power-law profile
(middle), and extended power-law profile (right), respectively.}
\end{figure*}

Let us start our analysis with the standard polynomial cosmographic
reconstruction and consider three lens mass density profiles: SIS,
power-law model, and extended power-law model. For the simplest SIS
model, the numerical and graphical results are respectively
presented in Table 1 and Fig.~7, with the best-fitted values for the
parameters: $\Omega_k=0.002\pm0.035$ and $f_E=1.000\pm0.002$. On the
one hand, one may clearly see the degeneracy between the cosmic
curvature and the lens model parameters, a tendency revealed and
extensively studied in the previous works \citep{Zhou20}. The
best-fitted value of $f_E$ is exactly what one could expect knowing
how the SGL data were simulated, i.e. the SIS velocity dispersion is
equal (up to some noise added) to the observed velocity dispersion
reported in mock catalog. This supports reliability of our
procedure. On the other hand, a spatially flat universe is supported
at much higher confidence levels ($\Delta \Omega_k \sim 10^{-2}$),
compared with the previous results obtained in
\citet{Xia2017,Qi2019} by applying the above procedure to different
available SGL subsamples. In the framework of power-law mass density
profile, one can derive constraint on the cosmic curvature as
$\Omega_k=-0.007\pm0.029$, with the best-fitted lens parameter and
the corresponding 1$\sigma$ uncertainty: $\gamma=2.000\pm0.012$. In
addition, it is worth noting that when the fractional uncertainty of
the Einstein radius is reduced to the level of 3\% (with HST
imaging), the resulting constraint on the cosmic curvature becomes
$\Delta \Omega_k=0.028$. Therefore, the estimation of the spatial
curvature is more sensitive to the measurements of lens velocity
dispersions, which indicates the importance of deriving additional
spectroscopic information for individual lenses. In the case of
extended power-law lens model, we get the weakest fits on the cosmic
curvature in the three types of lens models, with the best-fit value
and the marginalized 1$\sigma$ uncertainty $\Omega_k=0.003\pm0.045$.
Whereas, our analysis also yield improved constraints on the the
total-density and luminosity density profiles,
$\alpha=2.000\pm0.014$ and $\delta=2.171\pm0.035$. Compared with the
profile of the total mass, the density of luminous baryoic mass has
exhibited slight different distribution in early-type galaxies, i.e,
the stellar mass profile in the inner region of massive lensing
galaxies could fall off steeply than that of the total mass. Such
tendency, which has been revealed and studied in detail in
\citet{Cao2016}, might helpfully contribute to the understanding of
the presence of dark matter, which is differently spatially extended
than luminous baryons in early-type galaxies. More importantly,
besides the different degree of degeneracy between the lens model
parameters, our analysis also reveals the strong correlation between
$\Omega_k$ and the parameters characterizing the lens mass profiles.
Therefore, the large covariances of $\Omega_k$ with the power-law
parameters seen in Fig.~6 motivates the future use of auxiliary data
to improve constraints on the galaxy structure parameters. Now, the
question is: What is the average $\alpha$, $\delta$ and their
intrinsic scatter for the overall population of early-type galaxies?
One can use high-cadence, high-resolution and multi-filter imaging
of the resolved lensed images, to put accurate constrains on the
density profiles of galaxies
\citep{Suyu06,Vegetti10,Collett14,Wong15}, with the newly developed
state-of-the-art lens modeling techniques and kinematic modeling
methods \citep{Suyu12}. More specifically, the joint lensing and
dynamical studies of the SL2S lens sample have demonstrated that the
precision of 5\% could be obtained for the total-mass density slope
inside the Einstein radius \citep{Ruff2011,Sonnenfeld13} \footnote{
Note that the constraints on the mass density slope could be
improved to the level of 1\%, with precise time delay measurements
for the quasar-galaxy strong lensing systems \citep{Wucknitz04}.}
Hence, the LSST lenses should be technically supported by dedicated
follow-up imaging of the lensed images, possibly performed with more
frequent visits on Hubble telescope and smaller ground-based
telescopes. Meanwhile, observations of the lens galaxy spectra are
also needed in order to obtain the kinematic velocity dispersions,
which could be satisfied by Adaptive optics (AO) IFU spectroscopy on
8-40m-class telescopes. Other possible solutions to this issue can
simultaneously satisfy all of these needs, focusing on the
combination of AO imaging with slit spectroscopy \citep{Hlozek19}.

Another important issue is the choice of the $D_L(z)$ function
reconstructed from current quasar sample that served for the
$\Omega_k$ estimation. Therefore, we perform a similar analysis with
the logarithmic polynomial reconstruction and obtained the
constraints in the parameter space of $\Omega_k$ and ($f_E, \gamma,
\alpha, \delta$) for three cases of mass density profiles. The
results are also shown in Fig.~6 and Table 1. Comparing constraints
based on the two different reconstructions, we see that confidence
regions of different parameters (cosmic curvature and lens model
parameters) are well overlapped with each other; hence our results
and discussions presented above are robust. The strong degeneracies
between the cosmic curvature parameter and the lens model parameters
are also present as illustrated in Fig.~8. More interestingly,
compared with the standard polynomial reconstruction, the advantage
of the logarithmic polynomial reconstruction is that it could
provide more stringent constraints on the cosmic curvature:
$\Omega_k=-0.001\pm0.030$, $\Omega_k=-0.007\pm0.016$ and
$\Omega_k=0.002\pm0.031$, respectively in the framework of three
lens mass density profiles (SIS model, power-law spherical model,
and extended power-law model). Our results indicate that logarithmic
parametrization is a more reasonable approximation of theoretical
values up to high redshift. Such findings, which highlight the
importance of choosing a reliable $D_L(z)$ parametrization to better
investigate the spatial properties in the early universe, have also
been noted and discussed in the previous works
\citep{Risaliti2018,Melia19}. It should be noted that, even though
we focus on the simulated data of SGL systems trying to assess the
performance of the method in the future, the reconstructed distances
are obtained from the real data. Hence, the best-fitted values of
$\Omega_k$ somehow reflect what is supported by the observational
data.

Finally, an accurate reconstruction of cosmic curvature with
redshift can greatly contribute to our understanding of the
inflation models and fundamental physics. In order to address this
issue, we divide the full SGL sample into five groups with $\Delta
z_s=1.0$ (based on the source redshifts) and obtain the constraints
on $\Omega_k$ in the framework of SIS model. The first subsample has
400 SGL with source redshifts $z_s<1.0$, the second subsample has
2000 SGL with $1.0<z_s<2.0$, the third subsample has 1800 SGL with
$2.0<z_s<3.0$, the fourth subsample has 600 SGL with $3.0<z_s<4.0$,
and the fifth subsample contains 200 SGL with source redshifts
$4.0<z_s<5.0$. The corresponding results are shown in Fig.~7, with
the $d(z)$ function (polynomial and log-ploynomial parameterized
distance) reconstructed from the full quasar sample. Compared with
the previous analysis performed to test cosmic curvature with
different tests involving other popular astrophysical probes
including SNe Ia \citep{Xia2017} and compact radio quasars
\citep{Qi2019}, it is suggested that our technique, i.e., using
luminosity distance of quasars directly derived from the non-linear
relation between X-ray and UV luminosities, will considerably
improve such direct measurement of the spatial curvature in the
early universe ($z\sim5.0$).

\begin{figure}
\begin{center}
\includegraphics[width=0.95\linewidth]{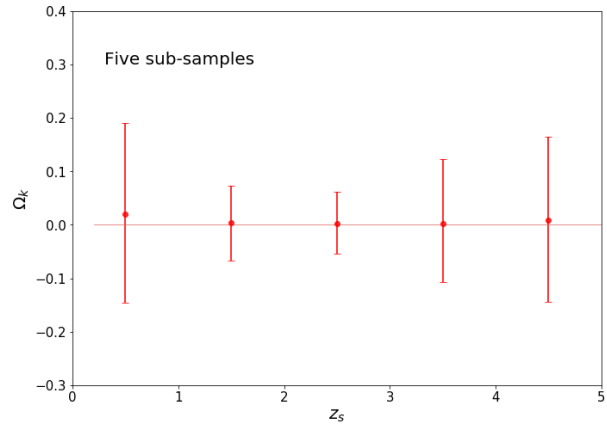}
\end{center}
\caption{Determination of cosmic curvature with five subsamples
$0<z<1.0$, $1.0<z<2.0$, $2.0<z<3.0$, $3.0<z<4.0$ and $4.0<z<5.0$
based on the source redshifts of SGL sample characterized by the SIS
lens model.}
\end{figure}

\section{Conclusions}

In this paper, we re-estimate which precision can be achieved for
the cosmic curvature in the near future, on the basis of the
distance sum rule in the well-known FLRW metric \citep{Rasanen2015}.
For the purpose, we focus on the simulated data of SGL systems
expected to be detected by LSST, combined with the recently
assembled catalog of 1598 high-quality quasars calibrated as
standard candles. It is demonstrated that in the framework of such
cosmological-model-independent way, the quasars have better coverage
of redshift in SGL systems at high redshifts, which makes it
possible to study the spatial properties in the early universe. Our
main conclusions are summarized as follows:

\begin{itemize}

\item Based on the future measurements of a particular well-selected subset of 5000 LSST lenses (with source redshifts $z\sim 5.0$),
the final results show that the the cosmic curvature could be
estimated with the precision of $\Delta \Omega_k \sim 10^{-2}$,
which is comparable to that derived from the Planck CMB power
spectra (TT, TE, EE+lowP) \citep{Planck Collaboration}. It should be
pointed out, even though the simulated data of SGL systems are used
in our analysis , the reconstructed distances are obtained from the
currently compiled quasar sample. In particular, no assumption on
the cosmic curvature is made the simulation of the LSST lens sample,
i.e., two SGL samples using the standard polynomial and logarithmic
polynomial cosmographic reconstructions. Therefore, the best-fitted
values of $\Omega_k$ somehow reflect what is supported by the real
observational data.

\item In our analysis, three types of models, which has been extensively investigated in
the literature is considered to describe the structure of the lens.
Specially, one may obtain the most stringent fits on the cosmic
curvature in the power-law lens model, while our $\Omega_k$
estimation will be strongly affected by the complicated extended
power law model (considering the possible difference between the
luminosity density profile ($\nu(r)\sim r^{-\delta}$) and the
total-mass). Furthermore, our analysis also reveals the strong
correlation between the cosmic curvature ($\Omega_k$) and parameters
characterizing the mass profile of lens galaxies ($f_E$, $\gamma$,
and $\alpha$, $\delta$), which motivates the future investigation of
lens density profiles through the combination of state-of-the-art
lens modeling techniques and kinematic modeling methods
\citep{Suyu12}. There are several sources of systematics that remain
to be discussed and addressed in the future analysis. The first one
is related to the galaxy structure parameters, especially those
characterizing the stellar distribution in the lensing galaxies. In
this paper, we adopted a power-law profile in the spherical Jeans
equation, with the aim of connecting the observed velocity
dispersion to the dynamical mass. However, many modern lens models
have considered a two-component model that is the sum of a
Sersic-like profile (fit to the stellar light distribution) and a
NFW profile (fit to the dark matter distribution) \citep{Navarro97}.
The luminosity distribution of spherical galaxies could also be well
described by the well-known Hernquist profile, whose behavior
follows an inner slope of $r^{-1}$ at small radii and $r^{-4}$ at
large radii \citep{Hernquist90}. Enlightened by the most recent
studies trying to quantify how cosmological constraints are altered
by different luminosity density profiles \citep{Ma19}, such effect
will contribute to the scatter in our cosmic-curvature test. This
also highlights the importance of auxiliary data in improving
constraints on the luminosity density profile, i.e., more
high-quality integral field unit (IFU) data are needed to further
improve the method in view of upcoming surveys \citep{Barnabe13}.

\item Our results indicate that, properly calibrated UV - X-ray
relation in quasars has a great potential of becoming an important
and precise distance estimator in cosmology. Based on the two
cosmographic reconstructions of $D_L(z)$ function, our findings also
highlight the importance of choosing a reliable reconstruction
schemes in order to better investigate the nature of space-time
geometry at high redshifts. This conclusion is also confirmed by the
the reconstruction of cosmic curvature with the source redshift
$z_s$, with accurate observations and spectral characterization of
quasars observed by SDSS \citep{Shen2011,Paris2017} and XMM
\citep{Rosen2016}. Finally, this paper seeks to highlights the
potential of LSST, which is expected to find extraordinary numbers
of new transients every night \citep{Smith19}. For instance, one
should recall that other promising settings for SGL systems have
been proposed, for example, galactic-scale strong gravitational
lensing systems with Type Ia supernovae \citep{Goobar17,Cao18,
Cao19c} and gravitational waves (GWs) as background sources
\citep{Liao2017b,Cao19a,Qi2019b,Qi2019c}. Benefit from LSST's
wide-field of view and sensitivity, these upcoming improvements on
the precision of cosmic curvature estimation will be very helpful
for revealing the physical mechanism of cosmic acceleration, or the
nature of cosmic origins.

\end{itemize}

\section*{Acknowledgments}

We are grateful to the referee for useful comments, which allowed to
improve our paper substantially. This work was supported by National
Key R\&D Program of China No. 2017YFA0402600; the National Natural
Science Foundation of China under Grants Nos. 11690023, and
11633001; Beijing Talents Fund of Organization Department of Beijing
Municipal Committee of the CPC; the Fundamental Research Funds for
the Central Universities and Scientific Research Foundation of
Beijing Normal University; and the Opening Project of Key Laboratory
of Computational Astrophysics, National Astronomical Observatories,
Chinese Academy of Sciences. M.B. was supported by the Key Foreign
Expert Program for the Central Universities No. X2018002. This work
was performed in part at Aspen Center for Physics, which is
supported by National Science Foundation grant PHY-1607611. This
work was partially supported by a grant from the Simons Foundation.
M.B. is grateful for this support. He is also grateful for support
from Polish Ministry of Science and Higher Education through the
grant DIR/WK/2018/12.


\label{lastpage}

\end{document}